\documentclass{phep}       

\journal{PHEP}
\vol{xx}

\received{06 February 2025}
\published{xx March 2025}

\usepackage{amsmath}
\usepackage{amssymb}
\usepackage[hidelinks]{hyperref}
\usepackage{yhmath}

\def\be{\begin{equation}}
\def\ee{\end{equation}}
\def\bea{\begin{eqnarray}}
\def\eea{\end{eqnarray}}

\renewcommand{\thefootnote}{\fnsymbol{footnote}}

\newcommand{\nn}{\nonumber}

\def\unit{\leavevmode\hbox{\small1\kern-3.6pt\normalsize1}}

\def\gtwid{\mathrel{\raise.3ex\hbox{$>$\kern-.75em\lower1ex\hbox{$\sim$}}}}
\def\ltwid{\mathrel{\raise.3ex\hbox{$<$\kern-.75em\lower1ex\hbox{$\sim$}}}}
\def\gev{{\rm \, Ge\kern-0.125em V}}
\def\tev{{\rm \, Te\kern-0.125em V}}

\def    \be            {\begin{equation}}
\def    \ee            {\end{equation}}
\def    \bea           {\begin{eqnarray}}
\def    \eea           {\end{eqnarray}}

\def\d{\delta}
\def\n{\nu}

\def\m{\mu}
\def\nn{\nonumber}
\def\d{\delta}

\def\s{\sigma}

\def\t{\theta}

 \normalsize

 \normalsize

\newcommand{\bmat}{\left(\begin{array}}
\newcommand{\emat}{\end{array}\right)}

\begin{document}

\title{Partial $\mu-\tau$ symmetry from kinetic normalization}

\author{N.~Chamoun,\auno{1,2}\footnotemark\ C.~Hamzaoui,\auno{3}  and M.~Toharia\auno{4}}
\address{$^1$Physics Department, HIAST, P.O. Box 31983, Damascus, Syria,}
\address{$^2$ CASP, Antioch Syrian University, Maaret Saidnaya, Damascus, Syria,}
\address{$^3$GPTP, D\'epartement des Sciences de la Terre et de L'Atmosph\`ere, \\
Universit\'e du Qu\'ebec \`a Montr\'eal, Case Postale 8888, \\
Succ. Centre-Ville, Montr\'eal, Qu\'ebec, Canada, H3C 3P8}
\address{$^4$Physics Department, Dawson College, 3040 Sherbrooke St., \\
Westmount, Qu\'ebec, Canada H3Z 1A4}

\begin{abstract}
We consider models with broken $\mu$-$\tau$ permutation symmetry through higher dimensional operators
  renormalizing the lepton kinetic terms in the action. We study
  the  consequences on the structure of the neutrino mass matrix and  find in particular that the allowed region for the lightest
  mass in the normal hierarchy plotted as a function of the atmospheric mixing element
  $|V_{\mu3}|$ is very restricted with the atmospheric mixing angle $\theta_{23}$ to lie
  in the second octant. On the other hand, the corresponding allowed region in the inverted hierarchy regime
  is less restrictive.
\end{abstract}

\maketitle


\renewcommand{\thefootnote}{\fnsymbol{footnote}}
\footnotetext{Corresponding author: nidalchamoun@gmail.com}

\section{Introduction}
Neutrinos offer a window into physics beyond standard model (BSM). Phenomenology suggested the existence of an approximate symmetry between the second and third lepton families  \cite{MUTAU}, but while this symmetry is known to be strongly broken in the charged lepton sector, its approximate phenomenological success in the neutrino sector may be due to corrections stemming from the charged lepton one \cite{Mohapatra0507312}. Actually, exact $\mu$-$\tau$ permutation symmetry ($M_{e\m}=M_{e\tau}$ and $M_{\m\m}=M_{\tau\tau}$) leads to vanishing smallest mixing angle $\t_{13}$ \cite{Harrison}, whence the need for corrections to remedy this experimentally excluded result. A slightly different variant called  $\mu$-$\tau$-reflection symmetry \cite{reflection}, imposing ($M_{e\m}=M_{e\tau}^*$ and $M_{\m\m}=M_{\tau\tau}^*$), allows for non-vanishing smallest mixing angle, but constrains much the phases in that Dirac phase should be right ($\d = \frac{\pi}{2}$) and the Majorana phases should be vanishing or right ($\eta, \xi =0, \frac{\pi}{2}$). In \cite{chamoun-hamzaoui-manuel}, we studied the phase breaking of the  $\mu$-$\tau$ symmetry ($|M_{e\m}|=|M_{e\tau}|$ and $|M_{\m\m}|=|M_{\tau\tau}|$), which can accommodate the experimental data but still with somewhat constraining predictions ($|V_{\mu i}| \approx |V_{\tau i}|$).

We study in this work the possibility where the breaking of the $\mu$-$\tau$ permutation symmetry comes from the non-canonical kinetic terms. To our knowledge, this has not been studied before.

\section{Notations}

Majorana neutrinos can, within seesaw scenarios, interpret the smallness of neutrino masses. Majorana mass term (${\n_L}^T M_\n {\cal C}{\n_L}$) (with $\cal C$ the charge conjugation matrix), implies a $12$-parameters complex symmetric matrix $M_\n$ which can be analysed into a $3$-parameters diagonal mass matrix $M_\n^{\mbox{\tiny diag.}}$ and a unitary $9$-parameters matrix $V$:
\be
M_\nu = V_\nu^* \left( \begin{matrix}m_1 & 0 & 0 \cr 0 & m_2 & 0 \cr 0 & 0
& m_3\end{matrix} \right) V_\nu^\dagger \nonumber
\ee
The $9$-parameters unitary matrix $V$ is analyzed into a product of three matrices:
\bea
V = P_\phi\;U_\d\;P^{\mbox{\tiny Maj.}}  &:& \nn
\\
P_\phi=\mbox{diag}\left(e^{i\phi_1},e^{i\phi_2},e^{i\phi_3}\right)&,&P^{\mbox{\tiny Maj.}} = \mbox{diag}\left(1, e^{i\eta},e^{i\xi}\right),
\nn
\eea
where the nonphysical phases ($\phi_{1,2,3}$) can be absorbed by redefining the charged lepton fields, so are non physical.

Working in the flavor basis, where charged lepton mass matrix is chosen -without loss of generality- to be diagonal, the lepton mixing entirely comes from the neutrino sector, and the measurable mixing matrix is parameterized as follows.
\bea
U_{\mbox{\tiny PMNS}} &=& U_\d\;P^{\mbox{\tiny Maj.}} \nn \eea
\bea
\label{pdg}
&\footnotesize{
 U_\d^{{\mbox{\tiny PDG}}}=\left ( \begin{array}{ccc} c_{12}\, c_{13}  & s_{12}\, c_{13} & s_{13} e^{-i\delta} \\ - s_{12}\, c_{23}- c_{12}\, s_{23}
\,s_{13} \, e^{i\delta}   &  c_{12}\, c_{23}\,- s_{12}\, s_{23}\, s_{13} e^{i\delta}
& s_{23}\, c_{13}\, \\ s_{12}s_{23}- c_{12}\, c_{23}\, s_{13} e^{i\delta}   &  - c_{12}\, s_{23}\, - s_{12}\, c_{23}\, s_{13} e^{i\delta}
 & c_{23}\, c_{13} \end{array} \right )
,} \nn
\eea

However, we prefer to work at the mixing matrix level, that we denote $V_{PMNS}$, and which can be parameterized with the $6$ degrees of freedom ($|V_{e2}|, |V_{e3}|, |V_{\mu 3}|, \d, \eta, \xi$) as follows:
\begin{eqnarray}
V_{PMNS}= \left(\begin{array}{ccc} |V_{e1}| & |V_{e2}|e^{i\eta} & |V_{e3}|e^{i(\xi-\delta)} \\
V_{\mu1} & V_{\mu2} & -|V_{\mu3}|e^{i\xi} \\ V_{\tau1} & V_{\tau2} &
|V_{\tau3}|e^{i\xi} \end{array}\right)
\label{vpmns} \nn
\end{eqnarray}
 Note that the (-) sign in the entry (2,3) w.r.t. PDG parametrization, means the second nonphysical phase in the two parameterizations are shifted by $\pi$.

\section{Experimental Data and Mass Hierarchy}
 The neutrino mass spectrum can be classified into two patterns:
 \begin{itemize}
\item Normal Hierarchy (NH) ($m_1$ lightest):
  $m_2=\sqrt{\Delta^2_{sol}+m_1^2}$\   and\ $m_3=\sqrt{\Delta^2_{atm} + m_1^2}$
\item Inverted Hierarchy (IH)  ($m_3$ lightest):
$m_1=\sqrt{\Delta^2_{atm} + m_3^2}$ \
  and\ $m_2=\sqrt{\Delta^2_{atm}+\Delta^2_{sol}+m_3^2}$.
\end{itemize}

The latest neutrino mixing global best fits \cite{data} is summarized in Table \ref{experimentalData}. Note that the constraints differ in the two hierarchies.

\begin{table}[h!]
\tbl{Latest Neutrino Mixing global best fits (1-$\s$ with SK atmospheric data).\label{experimentalData}}{%
\hspace{-1cm}
\begin{tabular}{c||c|c|}
\toprule
  &Normal Hierarchy&Inverted Hierarchy\\
\colrule
$|V^{exp}_{e3}|^2$  & $0.02215^{+0.00056}_{-0.00058}$  & $0.02231^{+0.00056}_{-0.00056}$  \\
\hline
$|V^{exp}_{e2}|^2$ & $0.30785^{+0.01200}_{-0.01100}$  & $ 0.30785^{+0.01200}_{-0.01100} $ \\
\hline
$|V^{exp}_{\mu 3}|^2$  & $0.46977^{+0.01700}_{-0.01300}$ & $0.54973^{+0.01201}_{-0.01101}$ \\
\hline
$\delta/\pi$ & $1.17778^{+0.14444}_{-0.22778}$ &  $1.52222^{+0.12222}_{-0.13889} $ \\
\hline
$\Delta^2_{sol}$ & $\left(7.49^{+0.19}_{-0.19}\right)\times 10^{-5}
    {\rm eV^2}$ &   $ \left(7.49^{+0.19}_{-0.19}\right)\times 10^{-5}
    {\rm eV^2}$  \\
\hline
$\Delta^2_{atm}$  & $\left(2.513^{+0.021}_{-0.019}  \right) \times 10^{-3} {\rm
  eV^2}$  &   $\left(-2.484^{+0.020}_{-0.020} \right) \times 10^{-3} {\rm eV^2} $ \\
\botrule
\end{tabular}}
\end{table}

\section{Texture}
We study a texture implementing an approximate $\mu$-$\tau$-symmetry, where the breaking is transmitted only via the
kinetic terms of the lepton $SU(2)$-doublets $L_\mu$ and $L_\tau$
through higher order operators.

Actually, as a toy model, one can assume the addition of three SM singlet scalar flavon
fields $\Phi_\alpha$ ($\alpha=e, \mu,\tau$), of physical dimension $n_\alpha$, coupled to the kinetic term through coupling coefficients $f_{\beta \gamma \alpha}$ inversely proportional to $\Lambda^{n_\alpha}$ with $\Lambda$ some BSM high energy scale:
\bea {\mathcal{L}}_{\mbox{\tiny kin.}} & \ni & f_{\beta \gamma \alpha} \Phi_\alpha \bar{L}_\beta \partial\!\!\!/ L_\gamma \eea
then, with plausible symmetry assumptions leading to the ansatz:
\bea\label{ansatz} f_{\beta \gamma \alpha} \neq 0 &\Rightarrow & \beta = \gamma = \alpha , \eea
we see that when the flavon field $\Phi_\alpha$ develops a VEV, denoted by $<\phi_\alpha>^{n_\alpha}$, with $\phi_\alpha$ of physical dimension 1, leading to flavor spontaneous symmetry breaking (SSB),
the modification of the kinetic
terms can be parameterized as $(1+ \frac{<\phi_\alpha>^{n_\alpha}}{\Lambda^{n_\alpha}})
\bar{L}_\alpha \partial\!\!\!/ L_\alpha$.

We need to canonically normalize the kinetic terms by a non-unitary redefinition of fields $L_\alpha \to \widetilde{L_\alpha}={L_\alpha} {\sqrt{1+
\frac{<\phi_\alpha>^{n_\alpha}}{\Lambda^{n_\alpha}}}}$. Consequently, the entry $M_{\alpha \beta}$ of the neutrino mass matrix transforms into:
\bea M_{\alpha \beta}  &\rightarrow& \widetilde{M_{\alpha \beta}} = \frac{M_{\alpha \beta}}{{\sqrt{1+
\frac{<\phi_\alpha>^{n_\alpha}}{\Lambda^{n_\alpha}}}}   {\sqrt{1+
\frac{<\phi_\beta>^{n_\beta}}{\Lambda^{n_\beta}}}}  } \eea
However, since our starting point was an exact $\mu$-$\tau$ symmetry, so one can rewrite the neutrino mass matrix, dropping the tilde, in the form:

\begin{eqnarray}
M_{\nu}= \left(\begin{array}{ccc} M_{ee} & M_{e \mu} & k M_{e \mu} \\
M_{e \mu} & M_{\mu \mu} & M_{\mu \tau} \\
k M_{e \mu} & M_{\mu \tau} & k^2 M_{\mu \mu} \end{array}\right) \label{permutationmatrix} \nn
\end{eqnarray}
where
$k =\left({\sqrt{1+\frac{<\phi_\tau>^{n_\tau}}{\Lambda^{n_\tau}}}}\right)/\left({\sqrt{1+
\frac{<\phi_\mu>^{n_\mu}}{\Lambda^{n_\mu}}}}\right)$. Note that $k$ is complex in
general, if the flavon VEVs are complex. We thus have two complex
constraints among the elements of the neutrino mass matrix (or four
real-constraints) \begin{eqnarray}
M_{e \tau} = kM_{e \mu} &,&
M_{\tau \tau} = k^2 M_{\mu \mu} \label{constraints}
\end{eqnarray}

\section{Discussion}
Absorbing the unphysical phases in the $M_\n$-entries (to be called $M_{ij}$ ($i,j=1,2,3$)), one can express the latter entries in terms of physical measurable quantities:
 \begin{eqnarray}
\label{M11} e^{-2i\phi_1} M_{ee}= M_{11} &=& m_1 V_{e1}^2+m_2 V_{e2}^2+ m_3 V_{e3}^2 \nn \\
 \label{M22} e^{-2i\phi_2} M_{\mu\mu}= M_{22} &=&  m_1 V_{\mu1}^2+ m_2 V_{\mu2}^2+ m_3 V_{\mu3}^2 \label{M22} \nn\\
\label{M33}  e^{-2i\phi_3} M_{\tau \tau}= M_{33} &=&  m_1 V_{\tau1}^2+ m_2 V_{\tau2}^2+ m_3 V_{\tau3}^2 \label{M33} \nn\\
\label {M12} e^{-i(\phi_1+\phi_2)} M_{e\mu}=M_{12} &=&  m_1 V_{e1}V_{\mu1}+ m_2 V_{e2}V_{\mu2}+ m_3 V_{e3}V_{\mu3} \label{M12} \nn\\
\label{M13} e^{-i(\phi_1+\phi_3)} M_{e\tau}= M_{13} &=&  m_1 V_{e1}V_{\tau1}+ m_2 V_{e2}V_{\tau2}+ m_3 V_{e3}V_{\tau3} \label{M13} \nn\\
\label{M23} e^{-i(\phi_2+\phi_3)} M_{\mu \tau}= M_{23} &=&  m_1 V_{\mu1}V_{\tau1}+ m_2 V_{\mu2}V_{\tau2}+ m_3 V_{\mu3}V_{\tau3} \label{M23} \nn\\
\label{Mij}
\end{eqnarray}
 Eliminating $k$ from Eq. \ref{constraints}, and using Eqs. \ref{Mij}, we get the complex relation
\begin{eqnarray}
M_{13}^2M_{22}=M_{12}^2M_{33} \label{constraint}
\end{eqnarray}
which is independent of the parameter $k$ and its phase, as well as of the unphysical phases $\phi_i$. From it we
can extract the dependance of $|V_{\mu3}|$ with respect to the rest of
$V_{\tiny PMNS}$ entries ($|V_{e2}|$, $|V_{e3}|$, $\delta$,
$\eta$ and $\xi$), as well as the neutrino mass parameters.
A second constraint among physical parameters (independent of $|V_{\mu
  3}|$) is also derived to further constrain the physical
parameters of the model.

\begin{figure}[h]
  \center
  \includegraphics[height=7cm,width=8cm]{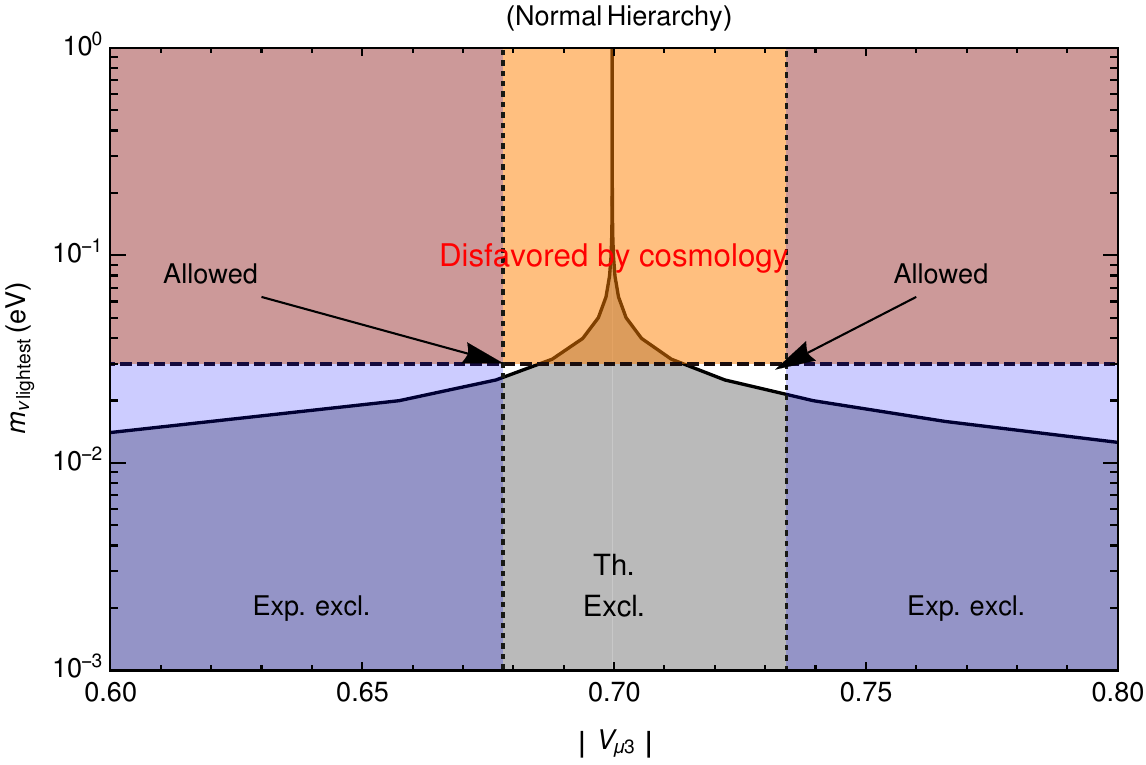}
  \caption{Lightest mass as a function of $|V_{\mu3}|$ in the NH ordering.}
\label{fig:etaxiN}
\end{figure}

Assuming NH ordering, Fig. \ref{fig:etaxiN} shows $m_1$ versus
$|V_{\mu 3}|$. The horizontal dotted line represents an upperbound on
the lightest neutrino mass coming from cosmological considerations,
whereas the two vertical dotted lines represent the limits from Table
\ref{experimentalData}. The solid line represents the exact numerical results
of the texture predictions from Eq.~(\ref{constraint}) which exclude the
region beneath it. We see that most of the parameter space is
disallowed, while the two ``white" small regions are allowed
experimentally, where the larger one points to $|V_{\mu 3}|$ larger
than $0.7$, which translates to $\t_{23}$ in the second octant.

\begin{figure}[h]
  \center
  \includegraphics[height=7cm,width=8cm]{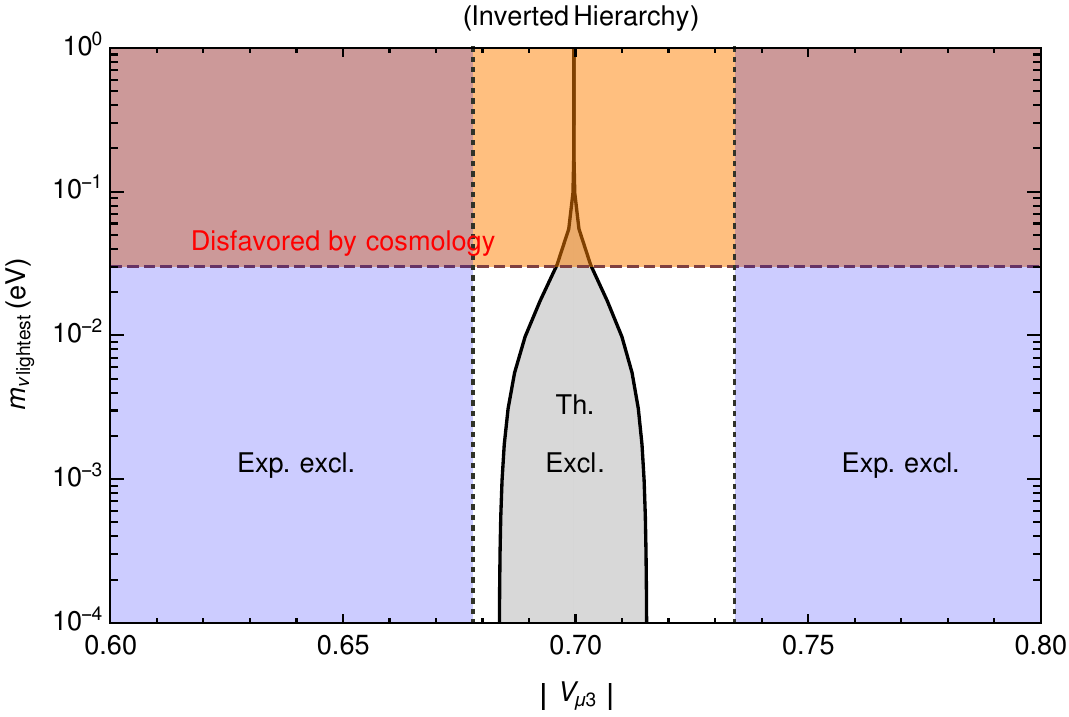}
  \caption{Lightest mass as a function of $|V_{\mu3}|$ in the IH ordering.}
\label{fig:etaxiI}
\end{figure}

On the other hand, Fig. \ref{fig:etaxiI} represents the corresponding
results assuming IH ordering and $m_3$ on the ordinate axis. We see
here that the allowed region is larger than in the NH one, but again
the region of $\t_{23}$ in upper octant is still bigger.

It is possible to derive exact relations from the two constraints from
Eq~.(\ref{constraint}) however they do not shed excessive light to the
numerical results shown in Figures \ref{fig:etaxiN} and \ref{fig:etaxiI}.
On the other hand if we consider the limit in which
the lightest neutrino is massless and we treat perturbatively the
small mixing angle $|V_{e3}|$ we can obtain simple and transparent
expressions that confirm the numerical results.

In order to simplify the notation we will define the mass parameters
$x=\frac{\min{|m_1|,|m_3|}}{\sqrt{\Delta_{atm}^2}}$ and
$r=\frac{\Delta_{sol}^2}{\Delta_{atm}^2}=R|V_{e3}|^2$ which will encode the
physical neutrino mass dependance in the expressions.

We will also define the parameter $Z$
\bea \nn
Z &=&
\frac{|V_{\mu3}|^2-|V_{\tau3}|^2}{|V_{e1}||V_{e2}||V_{e3}||V_{\mu3}||V_{\tau3}|},
\eea
that we will treat analitycally instead of $|V_{\mu3}|$.
The inverse expression to recover $|V_{\mu3}|^2$ out of  the parameter
$Z$ is
 \bea \label{exactResult}
|V_{\mu3}|^2 &=&
\frac{(1-|V_{e3}|^2)}{2}+\frac{|V_{e1}||V_{e2}||V_{e3}|(1-|V_{e3}|^2)Z}{2\sqrt{4+|V_{e1}|^2|V_{e2}|^2|V_{e3}|^2Z^2}}
\eea
With all this we can derive the predictions of the model on the two
orderings for the neutrino masses in the limit of very light lightest
neutrino:

\begin{itemize}
\item  NH ordering with $m_1=0$:  The constraint equation from
  Eq.~(\ref{constraint}) leads to
 \begin{eqnarray}
Z \simeq \pm \frac{2}{\sqrt{r}|V_{e1}|^2|V_{e2}|^2}
\end{eqnarray}
This expression is independent of the physical phases from the
neutrino sector. Using the experimental values for  the masses and
mixings this result leads to a prediction for the value of
$|V_{\mu3}|$ such that $|V_{\mu3}|^2 \simeq 0.046$ for the negative solution and
$|V_{\mu3}|^2 \simeq 0.93$ for the positive solution.
Both results are extremely far away from the experimental value of
$|V_{\mu3}|$ in agreement  with the large exclusion region observed in
figure \ref{fig:etaxiN}, which grows as the neutrino mass $m_1$ decreases.

\item IH ordering with $m_3=0$:  From the constraint equation of
  Eq.~(\ref{constraint}), we obtain an approximate expression for the minimum value possible of
 Z, i.e.
\begin{eqnarray}
Z_{min} \simeq \pm \frac{r}{|V_{e3}|^2} 
\end{eqnarray}
From this expression we can extract the predicted limiting values of $|V_{\mu3}|$ that lie
around  $\frac{(1-|V_{e3}|^2)}{2}$, which are $|V_{\mu3}| \lesssim 0.683$
and $|V_{\mu3}| \gtrsim 0.715$  in perfect agreement with the numerical results shown in
figure \ref{fig:etaxiI}.

\end{itemize}





\section{Summary \& Conclusions}

We studied in this work the breaking of the exact mu-tau permutation symmetry
through the kinetic term only, under the assumption that the main
contribution to the breaking comes from higher dimensional operators containing kinetic
terms.  This possibility has not yet been investigated before, in that
usually the breaking is studied at the level of the mass terms. The
fact that the breaking vevs are different in the directions of $\mu$
and $\tau$, leads to interesting phenomenology. We found that $|V_{^\m
  3}| > 0.7$ in NH privileging $\t_{23} > 45^o$, with the
acceptable parameter space being very tight. The same results happen in
IH ordering, although the allowed parameter space is larger. A more
complete study of this framework including the possible breaking of the
Tri-Bimaximal scenario is in progress \cite{Chamoun-Hamzaoui-Toharia}



\begin{thebibliography}{99}

\bibitem{MUTAU}  R.~N.~Mohapatra and S.~Nussinov,
  Phys.\ Rev.\  D {\bf 60}, 013002 (1999);
  C.~S.~Lam, Phys.\ Lett.\  B {\bf 507}, 214 (2001).

 \bibitem{Mohapatra0507312}  R.N. Mohapatra, W. Rodejohann, Phys.\ Rev.\ D{\bf 72}, 053001 (2005)

\bibitem{Harrison}
   P.~F.~Harrison and W.~G.~Scott, Phys.\ Lett.\  B {\bf 547}, 219 (2002).

\bibitem{reflection}
W. Grimus and L. Lavoura, Phys. Lett. B {\bf 579}, 113 (2004).

\bibitem{chamoun-hamzaoui-manuel}
N. Chamoun, C. Hamzaoui, S. Lashin, S. Nasri, M. Toharia,
Phys. Rev. D {\bf 104}, 015004 (2021)

\bibitem{data} Esteban, I., Gonzalez-Garcia, M.C., Maltoni, M. et al. NuFit-6.0: updated global analysis of three-flavor neutrino oscillations. J. High Energ. Phys. 2024, 216 (2025).

\bibitem{Chamoun-Hamzaoui-Toharia}
N. Chamoun, C. Hamzaoui, M. Toharia, Work in Progress.



 \end{thebibliography}
\end{document}